\documentstyle[preprint,aps,eqsecnum]{revtex}

\begin{document}

\thispagestyle{empty}
\setcounter{page}{0}

\title{ CLASSICAL KINETICS OF HARD
THERMAL PHENOMENA IN HIGH TEMPERATURE QCD\footnotemark[1]}

\footnotetext[1]{\baselineskip=12pt
Talk given at the workshop THERMO 95, Dalian, P.R. China, August 1995.\\
MIT-CTP\# 2467}

\author{Cristina Manuel}

\address{Center for Theoretical Physics,
Laboratory for Nuclear Science, and Department of Physics\\
Massachusetts Institute of Technology, Cambridge,
Massachusetts 02139}

\maketitle

\thispagestyle{empty}
\setcounter{page}{0}

\begin{abstract}
\baselineskip=13pt
$\!\!$Classical transport theory for colored particles is reviewed
and used  to derive the hard thermal loops of QCD.
A perturbative study of   the non-Abelian transport
equations that preserves their gauge symmetry
is used to compute the induced color current  in
a hot quark-gluon plasma. From this approach the effective
action of hard thermal loops can be derived.
This derivation is more direct than alternative ones
based on perturbative quantum field theory, and shows that hard
thermal effects in hot QCD are essentially {\it classical}.
\end{abstract}

%\noindent
%PACS No: 12.38.Mh, 51.10.+y, 11.10.Wx, 11.15.Kc
%\hfill\break
%\hbox to \hsize{CTP\  \hfil August 1995}
%\vskip-12pt
%\eject

\baselineskip=15pt
\pagestyle{empty}
\section{INTRODUCTION}
\label{sec1}

The purpose of this talk is to give a brief account of  the connection
between the Hard Thermal Loops (HTL) of QCD and the classical transport theory
of the
quark-gluon plasma \cite{KLLM}.

Currently there is an increasing interest in studying the finite
temperature regime of QCD. This regime could be attained in some
astrophysical settings and it could be essential to understand
the early universe. The high temperature phase of QCD is going to be
explored experimentally in future heavy ion colliders, and it is the phase
that we are going to consider here.

By high temperature we mean that we are going to consider QCD in its
deconfined phase, so that quarks and gluons can be treated as individual
particles. We also mean that the temperature is the only relevant scale
present in the theory, since it is much bigger than all the masses of the
particles of QCD and therefore these can be neglected in this
approach. Then the thermal
energies of the particles are very big, and therefore the effects of the
interactions with the gauge fields are comparatively small, so that it is
expected than in this regime the perturbative approximation is valid.

As  is well known, the naive perturbative analysis of high temperature
QCD fails completely. This was realized when
physical quantities, such as the gluonic
damping rate, were found to be gauge dependent when  computed
following the standard rules of quantum field theories at finite
temperature. The connection between expanding in loops and
expanding in the coupling constant $g$ is broken in this regime.
 As  was realized by Braaten and Pisarski \cite{BP1}, as well as by Frenkel
 and Taylor \cite{FT},
there are one-loop corrections, the HTL, which are as important as tree
amplitudes, and therefore they have to be included consistently in
all the computations to non-trivial order in $g$.

The HTL of QCD were thoroughly studied since their discovery, and
it was found that they obey some simple set of rules.
 HTL arise in one-loop diagrams where the external momenta are {\it soft}
 ($\sim g T$), while the internal loop momentum is {\it hard} ($\sim T$).
For an $SU(N)$ theory HTL appear in all multigluon amplitudes,
 and they are always
 proportional to the Debye mass squared $m_D ^2 = g^2 T^2 (N + N_F/2 )/3$,
where
$N_F$ is the number of quark flavors.
It is also remarkable that HTL are UV-finite and that they are gauge invariant.
 HTL have generically a momentum  dependence  of the type
 $1/ Q \cdot p$, where $Q$ is a  light-like four vector.

We are going to consider here only the HTL of gluonic amplitudes.
There is an infinite set of those HTL that
 have to be resummed into non-local effective  vertices and propagators
to be able to compute consistently at first order in the coupling constant
$g$. It is convenient, therefore, to have an effective action of HTL.
 An important development in this direction was carried out by Taylor
and Wong \cite{TW}, who were able to give an explicit expression of
 the effective
action of HTL, $\Gamma_{HTL}$. Taylor and Wong wrote as an {\it Ansatz} for
that action
\begin{equation}
\Gamma_{HTL} ={m_D^2\over 2}\,\int d^4\!x\,A^a_0(x)A^a_0(x)
-\int {d\Omega\over (2\pi)^3} \,W(A_+)\ ,
\label{GammaH}
\end{equation}
where $A_+$ is the projection of the gauge field $A_{\mu}$ over the
 light-like four vector $Q=(1, {\bf q})$, and $d\Omega$ denotes integration
over all angular directions of the unit vector ${\bf q}$. Equation
 (\ref{GammaH})
just means that $\Gamma_{HTL}$ should contain a mass term for the static
color electric field that  accounts for Debye screening in the plasma,
plus an unknown functional $W (A_+)$. Demanding that $\Gamma_{HTL}$  be gauge
invariant leads to an equation for $W$, namely
\begin{equation}
\partial_+\,{\delta W(A_+)\over \delta A_+}
+g\,\left[ A_+,{\delta W(A_+)\over \delta A_+}\right]
=2\,\pi^2\,m_D^2\, {\partial\over\partial x^0} A_+\ .
\label{3.26}
\end{equation}

Taylor and Wong were able to solve the above equation and give
a closed expression for $\Gamma_{HTL}$. This equation
has been identified with another one appearing in a completely
different context, that of  Chern-Simons theory  in 3 dimensions
at zero temperature \cite{EN}. This identification has been used to
derive a non-Abelian generalization of the Kubo equation \cite{JN}.

 Other derivations of the effective action of HTL have been
given in the literature \cite{BI}, \cite{JLL}.
All these approaches to HTL
involve quantum field theory, requiring very long and complicated
computations (introduction of gauge fixing, ghosts, etc).
One question arises: is the heavy machinery of quantum field theory
 required to study
hard thermal effects in the quark-gluon plasma?
Hard thermal effects are due exclusively to {\it thermal} fluctuations,
and therefore we should be able to describe them within a classical
formalism and in a simpler and more transparent way.

Let us mention here that HTL are not found exclusively in QCD. They also
appear in other theories, such as QED.  In that case, the
HTL  in the vacuum polarization tensor could be obtained
just by using classical kinetic theory for a plasma of electrons
and ions \cite{Silin}. The same can be done for QCD.
 In the following sections we will show how  to derive
the HTL of QCD from the classical transport theory describing the
quark-gluon plasma.

\section{CLASSICAL TRANSPORT THEORY FOR A NON-ABELIAN
PLASMA}
\label{sec2}

The classical transport theory for the QCD plasma was developed
in~\cite{EH}, and here we will briefly review it. Consider a particle
bearing a non-Abelian $SU(N)$ color charge $Q^{a}, \ a=1,...,N^2-1$,
traversing a worldline $x^{\alpha}(\tau)$.  The Wong equations
\cite{Wong} describe the dynamical evolution of the
variables
 $x^{\mu}$, $p^{\mu}$ and
$Q^{a}$ (we neglect here the effect of  spin):
\begin{mathletters}
\label{wongeq}
\begin{eqnarray}
m\, {{d x^{\mu}}\over{d \tau}} & = & p^{\mu}
\ , \label{wongeqa} \\[2mm]
m\, {{d p^{\mu}}\over{d \tau}} & = & g\,Q^{a}F^{\mu\nu}_{a}p_{\nu}
\ ,\label{wongeqb}\\[2mm]
m\, {{d Q^{a}}\over{d \tau}} & = & - g\,
f^{abc}p^{\mu}A^{b}_{\mu}Q^{c}\ . \label{wongeqc}
\end{eqnarray}
\end{mathletters}
The main difference between the equations of electromagnetism and the
Wong equations, apart from their intrinsic non-Abelian structure, comes
from the fact that color charges precess in color space, and therefore
they are dynamical variables. Equation (\ref{wongeqc}) guarantees
 that the color current associated to each colored particle,
$j_{\mu} ^a (x) = g \int d \tau Q^a  p_{\mu} \delta^{(4)} ( x - x(\tau))$,
is covariantly conserved,
$ \left(D_{\mu} j^{\mu} \right)^a (x) = 0$, keeping therefore the
 consistency of the theory.

The usual $(x,p)$ phase-space is thus enlarged to $(x,p,Q)$ by
including  color degrees of freedom for colored particles.
Physical constraints are enforced by inserting delta-functions in
the phase-space volume element $dx\,dP\,dQ$. The momentum
measure
\begin{equation}
dP = {{d^{4}p}\over{(2\pi)^{3}}}\,\,2\,\theta(p_{0})\,\,
\delta(p^{2} - m^{2})
\label{measurep}
\end{equation}
guarantees positivity of the energy and on-shell evolution. The color
charge measure enforces the conservation of the group invariants,
{\it e.g.}, for $SU(3)$,
\begin{equation}
dQ = d^8 Q\,\, \delta(Q_{a}Q^{a} - q_{2})\,\,
\delta(d_{abc}Q^{a}Q^{b}Q^{c} - q_{3}) \ ,
\label{measureq}
\end{equation}
where the constants $q_{2}$ and $q_{3}$ fix the values of the
Casimirs and $d_{abc}$ are the totally symmetric group constants.
The color charges which now span the phase-space are dependent
variables. These can be formally related to a set of independent
phase-space Darboux variables \cite{KLLM}. For the sake of simplicity, we will
keep on using the standard color charges.

The one-particle distribution function $f(x,p,Q)$ denotes the
probability for finding the particle in the state $(x,p,Q)$.
In the collisionless case,  it evolves in
time via a transport equation
$ {{d f}\over{d \tau}} = 0$. Using the equations of motion~(\ref{wongeq}),
it becomes the Boltzmann equation
\begin{equation}
p^{\mu}\left[{{\partial}\over{\partial x^{\mu}}}
- g\, Q_{a}F^{a}_{\mu\nu}{{\partial}\over{\partial p_{\nu}}}
- g\, f_{abc}A^{b}_{\mu} Q^{c}{{\partial}\over{\partial Q_{a}}}
\right] f(x,p,Q) = 0 \ .
\label{boltzmann}
\end{equation}

A complete, self-consistent set of non-Abelian Vlasov equations for
the distribution function and the mean color field is obtained by
augmenting the Boltzmann equation with the Yang-Mills equations:
\begin{equation}
[D_\nu F^{\nu\mu}]^a(x) = J^{\mu\, a}(x) =  \sum_{\rm species}\
 \sum_{\rm helicities}\
j^{\mu\, a}(x)\ ,
\label{yangmills}
\end{equation}
where the  color current $j^{\mu\,a}(x)$
for each particle species is
computed from the corresponding distribution function as
\begin{equation}
j^{\mu\,a} (x) = g\, \int dPdQ\ p^\mu Q^a f(x,p,Q) \ .
\label{cr5}
\end{equation}
It can be shown by using the
Boltzmann equation  that the color current is covariantly conserved,
$ \left(D_{\mu} j^{\mu} \right)^a (x) = 0 $.

 The Wong equations (\ref{wongeq})
are  invariant  under the finite gauge
transformations (in matrix notation)
\begin{equation}
\label{gaugetrsf}
\bar{x}^{\mu}=x^{\mu}\ , \qquad
\bar{p}^{\mu}= p^{\mu}\ , \qquad
\bar{Q} =  U \,Q \,U^{-1}\ , \qquad
{\bar A}_\mu = U\,A_\mu \,U^{-1}-{1\over g}\,U\,
{\partial\over \partial x_\mu}\,U^{-1}\ ,
\end{equation}
$\!\!$where $U(x)={\rm exp}[-g\,\varepsilon^a(x)\,t^a]$ is a group element.

It is easy to show that the Boltzmann equation (\ref{boltzmann}) is
 invariant under the above gauge transformation if
the distribution function behaves as an scalar
\begin{equation}
{\bar f}({\bar x},{\bar p},{\bar Q}) = f (x,p,Q).
\end{equation}
To check this statement it is important to note that under a gauge
transformation the derivatives appearing in the Boltzmann equation
(\ref{boltzmann}) transform as:
\begin{equation}
{\partial\over\partial x^\mu}=
{\partial\over\partial\bar{x}^\mu}
- 2 ~{\rm Tr}~ \Biggl([\ ({\partial\over\partial {\bar x}^\mu}U)
U^{-1}\ ,\  \bar{Q}\ ]
{\partial\over\partial\bar{Q}}\Biggr) \ , \qquad
{\partial\over\partial p^\mu}=
{\partial\over\partial\bar{p}^\mu} \ , \qquad
{\partial\over\partial Q}=
U^{-1} {\partial\over\partial\bar{Q}}U
\label{eq:gauge2c}\ ,
\end{equation}
that is, they are not gauge invariant by themselves. Only
the specific combination of the spacial and color derivatives
that appears in (\ref{boltzmann}) which is
gauge invariant.

It is  also easy to show that the color current
(\ref{cr5}) transforms under (\ref{gaugetrsf}) as a gauge covariant vector:
\begin{equation}
{\bar j}^{\mu}({\bar x})=\int dP\,dQ\,p^\mu\,U\,Q\,
U^{-1}\,f(x,p,Q)=U\,j^{\mu}(x)\,U^{-1}\ .
\end{equation}
This is due to the gauge invariance of the phase-space measure and to the
transformation properties of $f$.

\section{EMERGENCE OF HARD THERMAL LOOPS}
\label{sec3}

We are now ready to use the formalism presented above to derive the
HTL of QCD. We consider  a hot, color-neutral quark-gluon plasma
close to equilibrium, so that the distribution function can be
expanded in powers of $g$:
\begin{equation}
f=f^{(0)}+gf^{(1)}+g^2f^{(2)}+...\ ,
\label{L1}
\end{equation}
where $f^{(0)}$ is the equilibrium distribution function in the
absence of a net color field, and is given, up to a normalization constant,
 by $n_{B,F}(p_0)=1/(e^{\beta
|p_0|}\mp 1)$, that is, it is  the bosonic/fermionic probability distribution.

 The Boltzmann equation (\ref{boltzmann}) for $f^{(1)}$ reduces to
\begin{equation}
p^{\mu} \left({\partial\over\partial x^{\mu}}-g\, f^{abc} A_{\mu}^b
Q_c {\partial\over\partial Q^a}\right)
f^{(1)}(x,p,Q) = p^{\mu} Q_a F_{\mu \nu}^a {\partial\over \partial
p_{\nu}} f^{(0)}(p_0)\ .
\label{L3}
\end{equation}
Notice that a complete linearization of the equation in $A_{\mu}$
would break gauge invariance. But notice as well that this
approximation tells us that $f^{(1)}$ also carries a $g$-dependence.
{}From (\ref{L3}) we can get the following equation that the total
 color current  density $J^{\mu} (x,p)$  has to obey \cite{KLLM}
\begin{equation}
[\,p \cdot D\,\, J^{\mu}(x,p)]^a = 2\,g^2\, p^{\mu} p^{\nu}
F_{\nu 0}^a {d \over dp_0}[N\, n_B(p_0)+N_F\, n_F(p_0)] \ .
\label{L5}
\end{equation}
Similar results have been obtained in~\cite{BI,JLL}, in a quantum
field theoretic setting.

{}From this equation it is possible to derive the effective action of
HTL, $\Gamma_{HTL}$. Necessary steps are: {\it i})  integrate
(\ref{L5}) over  $|{\bf p}|$ and $p_0$
using the massless limit of the momentum measure $dP$
(\ref{measurep}); {\it ii})   define a new current density
\begin{equation}
{\tilde {\cal J}}^{\mu}(x,v) = {\cal J}^{\mu}(x,v) + 2
\,\pi^2 \,m^2_D\, v^\mu A_0(x)\ ,
\label{3.13}
\end{equation}
where $v$ is the  light-like four vector, which in this case corresponds
to the four velocity vector of the colored particles of the plasma;
{\it iii})  assume that the color current ${\tilde J}^{\mu}$ can
 be derived from a generating functional as
\begin{equation}
{\tilde {\cal J}}^{\mu}(x,v)={{\delta W(A,v)}\over
{\delta A_\mu(x)}}\ .
\label{3.15}
\end{equation}
Under these assumptions  equation (\ref{L5}) becomes
exactly
 the equation  that expresses the gauge invariance
condition of $\Gamma_{HTL}$ (\ref{3.26}).
If we now define an
effective action $\Gamma$ that generates the color current, {\it i.e.},
$J^\mu(x)=-\frac{\delta\Gamma[A(x)]}{\delta A_\mu(x)}$,
then we are able to obtain an  expression for that action, which
coincides  exactly with $\Gamma_{HTL}$.

As a simple  application of the classical transport formalism presented
above, one can  solve  the  approximate Boltzmann equation (\ref{L3}) for
plane-wave excitations in the quark-gluon plasma.
In a plane-wave {\it Ansatz} in
which the vector gauge fields  only depend  on $x^\mu$ through the
combination $x \cdot k$, where $k^\mu = (\omega, {\bf k})$ is the
wave vector,  one can easily find the solution to the Boltzmann equation.
 The color current is given in this case by
\begin{equation}
J^{\mu}_a (x) =  m_D ^2 \int \frac {d \Omega} {4 \pi} \, v^{\mu}
\left (\omega \, \frac{ v \cdot A_a (x)} {v \cdot k}- A^0 _a (x) \right)
\ .
\label{curr}
\end{equation}

The  polarization tensor $\Pi^{\mu\nu}_{a b}$ can be computed from
(\ref{curr}) by using the relation
\begin{equation}
J^{\mu}_a (x) = \int d^4 y \, \Pi ^{\mu\nu} _{a b} (x-y)\,
A_{\nu} ^{b} (y) \ .
\end{equation}
It reads:
\begin{equation}
\Pi^{\mu\nu}_{ab} (\omega, {\bf k})=  m_D ^2 \left (-g^{\mu 0} g^{\nu 0} +
\omega \,  \int \frac{d \Omega}{4 \pi} \,
\frac {v^{\mu} v^{\nu}} {\omega - \bf {k} \cdot {\bf v} + i \epsilon} \,\
\right)
\delta_{ab} \ ,
\label{polar}
\end{equation}
where retarded boundary conditions have been imposed with the prescription
$\omega + i \epsilon$.
These results for the HTLs of the polarization tensor agree with those
obtained in the high temperature limit using quantum field theoretic
techniques.

\section{CONCLUSIONS}
\label{sec5}

We have shown how to compute the induced color current in a hot
quark-gluon plasma  using  classical transport theory.
This computation makes use of an approximation
that  respects the gauge symmetry of the transport equations.
By maintaining gauge invariance we have been able to extract from
this formalism the effective action for the infinite set of hard
thermal loops of QCD.

This derivation is remarkable because of its simplicity. It also
shows that hard thermal effects in a hot QCD plasma are classical.
That is, they are only due to thermal fluctuations and
 creation and annihilation processes play no role on them.

Classical transport theory
 can be positively
used to study hard thermal phenomena in the quark-gluon plasma.
It is a  very simple, direct and transparent formalism.
  It is obvious, however,
that it cannot give a complete description
of  next-to-leading order effects in high temperature QCD.

\vspace{15mm}
{\bf Acknowledgements:}
 This work  is supported by the
Ministerio de Educaci\'on y Ciencia, Spain, and by the
U.S.~Department of Energy (D.O.E.) under cooperative agreement
\#~DE-FC02-94ER40818.

\end{document}